\documentclass[twocolumn,superscriptaddress,floatfix]{revtex4-2}

\usepackage{lineno}
\usepackage{graphicx}
\usepackage{amssymb}
\usepackage{amsmath}
\usepackage{booktabs}
\usepackage{multirow}
\usepackage{url}

\usepackage[colorlinks=true,linkcolor=blue,breaklinks]{hyperref}

\begin{document}

\title[]{Restoring Original Signal From Pile-up Signal using Deep Learning}

\author{C.~H.~\surname{Kim}}
\affiliation{Department of Physics, Sungkyunkwan University, Suwon 16419, Republic of Korea}

\author{S.~\surname{Ahn}}
\affiliation{Center for Exotic Nuclear Studies, Institute for Basic Science (IBS), Daejeon 34126, Republic of Korea}

\author{K.~Y.~\surname{Chae}}
\email{kchae@skku.edu}
\thanks{Fax: +82-31-290-7055}
\affiliation{Department of Physics, Sungkyunkwan University, Suwon 16419, Republic of Korea}

\author{J.~Hooker}
\affiliation{Department of Physics $\&$ Astronomy, Texas A$\&$M University, College Station, TX 77843, USA}
\affiliation{Cyclotron Institute, Texas A$\&$M University, College Station, TX 77843, USA}

\author{G.~V.~Rogachev}
\affiliation{Department of Physics $\&$ Astronomy, Texas A$\&$M University, College Station, TX 77843, USA}
\affiliation{Cyclotron Institute, Texas A$\&$M University, College Station, TX 77843, USA}

\begin{abstract}

Pile-up signals are frequently produced in experimental physics. They create inaccurate physics data with high uncertainty and cause various problems. Therefore, the correction to pile-up signals is crucially required. In this study, we implemented a deep learning method to restore the original signals from the pile-up signals. We showed that a deep learning model could accurately reconstruct the original signal waveforms from the pile-up waveforms. By substituting the pile-up signals with the original signals predicted by the model, the energy and timing resolutions of the data are notably enhanced. The model implementation significantly improved the quality of the particle identification plot and particle tracks. This method is applicable to similar problems, such as separating multiple signals or correcting pile-up signals with other types of noises and backgrounds.

\end{abstract}

\date{\today}
\maketitle





\section{\label{sec:level1}Introduction}

Pile-up signals cause difficulties in experimental physics. In experiments with a high count rate or high background rate, the desired signals are constantly piled up with noises, background, or each other \cite{Z.Marshall2014, M.Hammad2019, M.Mohammadian-Behbahani2020}. Piling-up deforms spectra and information of original signals. Physics information such as height and timing can be distorted, leading to an inaccurate conclusion. Accurate corrections to pile-up signals are required to avoid this problem.

Deep learning methods have shown their innovative performance in various scientific research \cite{ImageClassification, B.Rem2019, H.Gabbard2022, A.Senior2020, J.Jumper2021, M.Raissi2019, C.Kim2023}. They are frequently more effective than conventional rule-based methods \cite{C.Zhang2016, C.Yang2017, P.Holl2019, L.Garcia2021, C.Kim2023}. The correction using the rule-based method, such as fitting, can be successful for simple pile-up signals. However, such a method could be inapplicable when it comes to the complicated shape of signals. On the other hand, deep learning methods are capable of handling such complications. Numerous studies have adopted deep learning for pulse shape analysis, mostly focusing on shape discrimination \cite{C.Zhang2016, C.Yang2017, P.Holl2019, L.Garcia2021, C.Kim2023}. They have demonstrated successful applications in discriminating types of particles or noises on various detector systems. For handling pile-up signals, Kafaee \textit{et al.} \cite{M.Kafaee2009} implemented a simple neural network that takes pile-up signals and outputs amplitudes and timing information. They presented an effective model with low errors.

A denoising autoencoder is one of the variations of autoencoders \cite{P.Vincent2008, P.Vincent2010}. While it can be utilized for general purposes of autoencoders, such as extracting valuable features, it is also effective to remove noises in data. A denoising autoencoder takes inputs with artificially added noises, such as Gaussian noises. It is trained to predict original inputs, which is an unsupervised way. The trained model can then restore the original data by simply observing the noisy data.

In this study, we restored the original signal waveforms from pile-up waveforms with noises in data obtained from the Texas Active Target (TexAT) detector system \cite{E.Koshchiy2020}. The TexAT is a time projection chamber with a Micromegas to reconstruct particle paths using timing information. The recorded energy and reconstructed particle path are distorted if a particle signal coincides with a background noise signal. The distortion results in low-quality data, such as low energy resolution and high errors in particle track fitting. We implemented the idea of denoising autoencoders with a 1-dimensional convolutional neural network to recover the original signals.

Section~\ref{sec:level2} presents the description of the experimental data used in this study. The data preparation for the deep learning model is described in Section~\ref{sec:level3}. The model construction details are presented in Section~\ref{sec:level4}, which is followed by a discussion on the model performance in Section~\ref{sec:level5}. Section~\ref{sec:level6} concludes this study and presents potential future applications.


\section{\label{sec:level2}Experimental Data from TexAT}

The TexAT has been used for various nuclear physics experiments, measuring nuclear reactions and decays \cite{J.Hooker2019, E.Koshchiy2020, J.Bishop2020, J.Zamora2021, J.Bishop2022}. The detailed descriptions of the TexAT and its typical data properties are presented in Refs. \cite{E.Koshchiy2020, C.Kim2023}. Here, we briefly present the main features related to this study.

\begin{figure}[!h]
\includegraphics[width=3.3in]{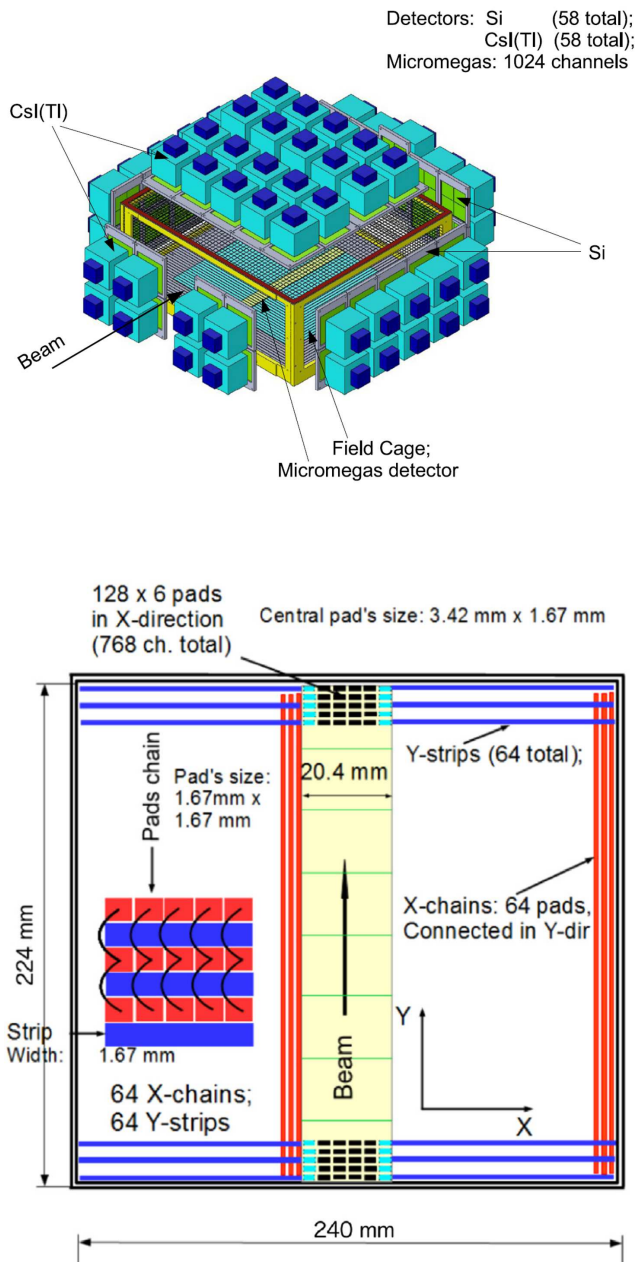}	
\caption{\label{fig:fig1}Schematic figures of the TexAT (top panel) and Micromegas readout board (bottom panel) \cite{E.Koshchiy2020}. The figures were taken from Ref. \cite{E.Koshchiy2020}.}
\end{figure}

A basic sketch of the TexAT detector system is shown in Fig.~\ref{fig:fig1}. The main components of the TexAT are a Micromegas, Si detectors, and CsI detectors. The chamber is filled with gas that plays both target and ionization medium roles. If the nuclear reactions or decays occur inside the chamber, their information is first measured by the Micromegas. The Micromegas is used to reconstruct particle tracks with two different regions: central and side regions. The central region of the Micromegas is composed of rectangular pads, and the side region is composed of strips and interconnected chains. For the side region, the particle tracks are reconstructed by matching the timings of the strips (y-axis) and chains (x-axis). The reconstructed particle tracks are analyzed to estimate angular distributions, reaction points, and so on. If the timing information of signals detected in the strips and chains is distorted, inaccurate particle paths will be reconstructed.

Particle identification (PID) can be performed with $\Delta E$ and $E$. A particle from a nuclear reaction or decay will typically lose some of its energy in the gas and the rest in the solid-state detectors. The energy loss in the gas detected by the Micromegas is used to estimate $\Delta E$, except for the particles that reach the CsI detectors. If the energy information of the signals detected in the Micromegas is distorted, it can create a PID plot with low energy resolution.

\begin{figure*}[!t]
\includegraphics[width=7in]{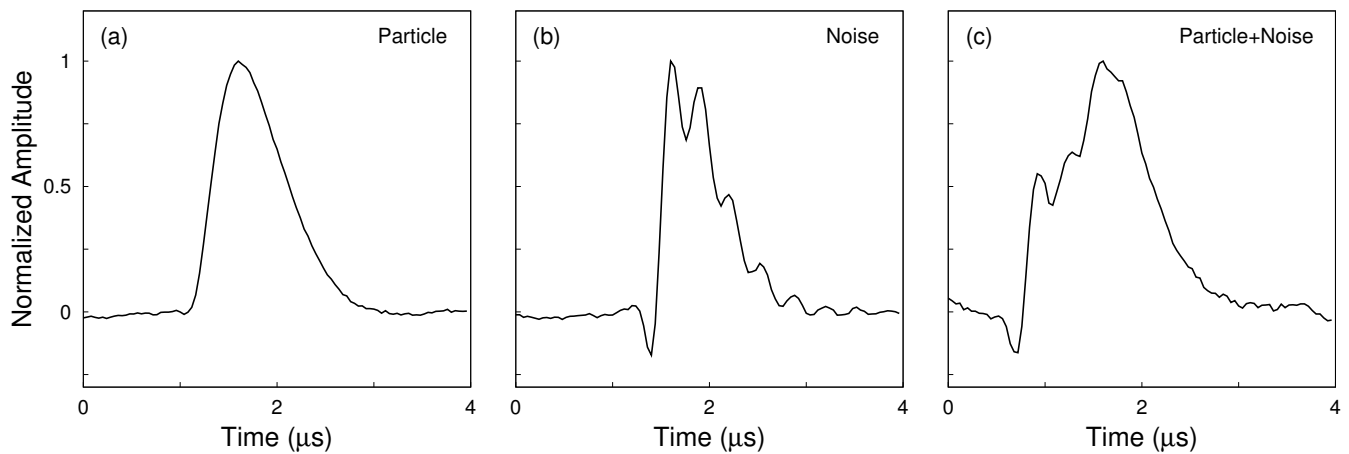}
\caption{\label{fig:fig2}Typical digital waveforms of particle, noise, and pile-up signals (particle+noise). Each signal is from a single channel of the Micromegas.}
\end{figure*}

The experimental data from the previous $^8B$+$p$ measurement was used for this study \cite{J.Hooker2019}. Detailed descriptions and results of the experiment can be found in Ref. \cite{J.Hooker2019}. The signals in this experiment were stored in 512 time buckets as digital waveforms with a bucket size of 40 ns. Unknown signals were frequently observed in the Micromegas detector. In this paper, we refer to the signals as noise signals. In the previous study of Kim \textit{et al.} \cite{C.Kim2023}, signals obtained in the experiments were classified into the particle, noise, and pile-up signals (particle+noise) using a deep classifier with high accuracy.

Typical waveforms of these signals are presented in Fig.~\ref{fig:fig2}. Fig.~\ref{fig:fig2} (a) and (b) show typical waveforms of particle and noise signals, respectively. The particle signals exhibit clean skewed bell-shaped waveforms. On the other hand, the noise signals have complicated waveform shapes with notable fluctuations. Fig.~\ref{fig:fig2} (c) shows the waveforms of the pile-up signals whose shapes appear to be the two waveforms linearly piled up. Note that the waveform shapes of noise signals or pile-up signals are inconsistent without a clear quantitative feature. For this reason, it is challenging to have an effective conventional rule-based method for recovering the waveforms of the original signals from those of the pile-up signals.

\section{\label{sec:level3}Data Preparation}

The data for machine learning is divided into three: training, validation, and test datasets. The training dataset is used for the model training. The validation dataset is utilized to assess the performance during hyperparameter tuning of the model, whereas the test dataset is utilized for the final assessment. Labeled data is required in supervised learning to learn the pattern between the inputs and labels. On the other hand, the machine is trained on untagged data for unsupervised learning.

\begin{figure*}[!t]
\includegraphics[width=7in]{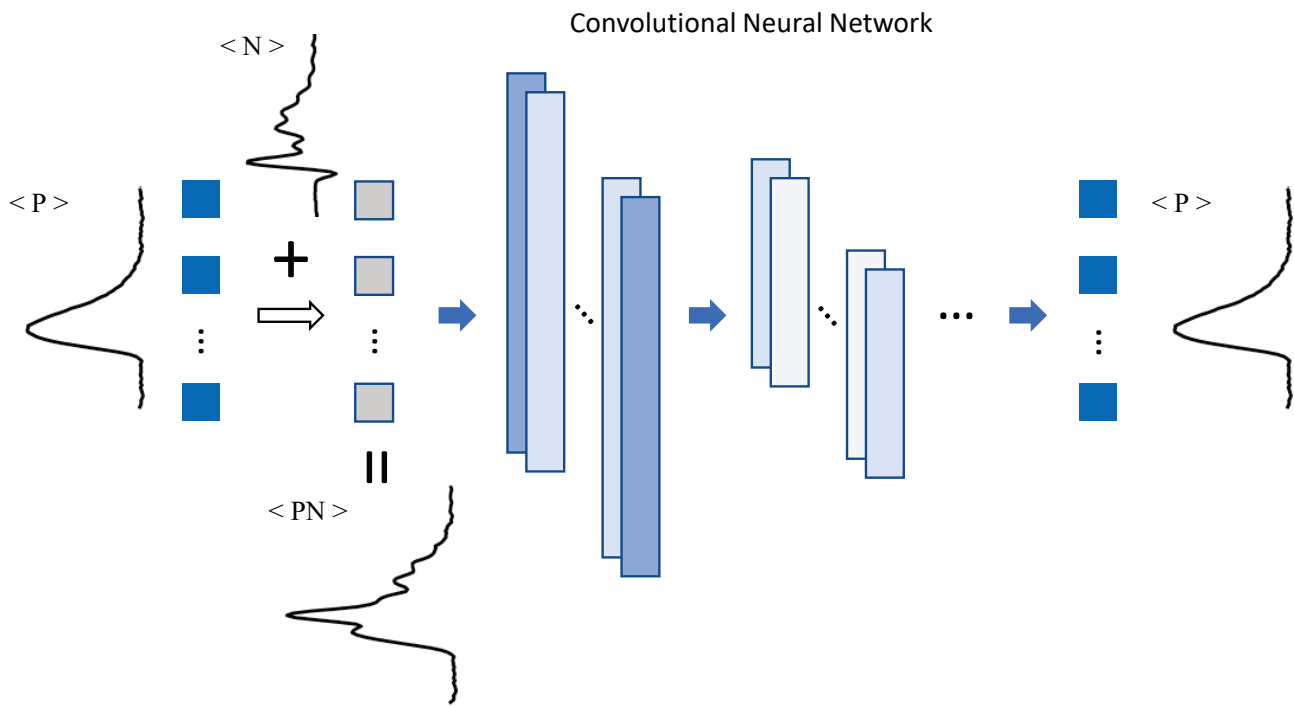}	
\caption{\label{fig:fig3}Sketch of denoising autoencoder in this study. $<$P$>$, $<$N$>$, and $<$PN$>$ represent particle, noise, and pile-up (particle+noise) signals, respectively. This process is equivalent to supervised learning with inputs and labels as pile-up and original particle signals, respectively. See text for more details.}
\end{figure*}

Training data for a typical denoising autoencoder does not include labels (unsupervised learning) \cite{P.Vincent2008, P.Vincent2010}. Random noises are added to the inputs before they are fed to the autoencoder. The autoencoder is then trained to return the original inputs. In this study, we used the idea of a denoising autoencoder to build a model that can recover the original signals from pile-up signals with noises. However, unlike the typical denoising autoencoder, the noise signals in the current data are not completely random noises but complicated noises created in a certain manner. Therefore, we prepared the data by the following procedure. We assume that the pile-up waveforms were produced by linearly adding particle and noise waveforms. This assumption was valid in the previous study of Kim \textit{et al.} \cite{C.Kim2023}. The inputs and outputs of our autoencoder are the particle waveforms. The input particle waveforms are artificially piled up with the noise waveforms before feeding them to the autoencoder. This process is summarized in Fig.~\ref{fig:fig3}. In practice, this is equivalent to supervised learning with inputs as pile-up signals and labels as particle signals.

\begin{figure}[!t]
\includegraphics[width=3.3in]{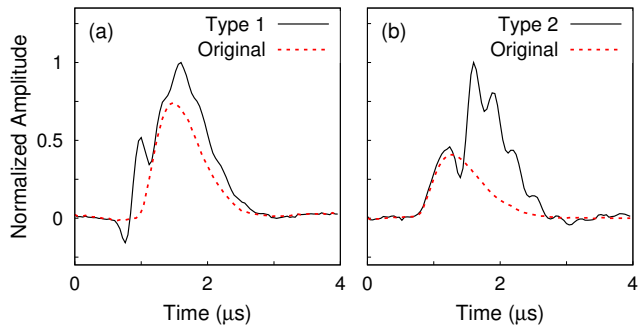}	
\caption{\label{fig:fig4}Input particle (original) and pile-up signals for model training. The pile-up signals are categorized into Types 1 and 2. See text for more details.}
\end{figure}

The manually labeled particle and noise signals from the previous study of Kim \textit{et al.} \cite{C.Kim2023} were used for the data preparation. The study used artificially combined particle and noise signals to increase the data sample size. The detailed labeling process can be found in the paper \cite{C.Kim2023}. Here, similar to Kim \textit{et al.} \cite{C.Kim2023}, the signals of two classes, particle, and noise, were linearly combined to obtain the input pile-up signals. Relative heights and positions of the particle and noise signals were determined in a random manner. The pile-up signals were categorized into two types: Type 1 if the particle signal is larger than the noise signal, and Type 2 if vice versa. Fig.~\ref{fig:fig4} shows typical waveforms of input particle (original) signals and the corresponding pile-up signals for each type. The Type 1 signal contains a relatively clear shape of the particle signal, whereas the Type 2 signal weakly exhibits the shape of the particle signal. At the evaluation time, we assess the model for each type to see if the performance of the model is biased to a particular type. Peaks of the pile-up signals were normalized to 1 and aligned at the 41st time bucket (bin). The length of waveforms was 100 buckets with a bin size of 40 ns. The total number of signals used in this study was about 40,000, with an equal number of Type 1 and 2 signals. The data was sliced into three: 60$\%$ of those for training, 20$\%$ for validation, and 20$\%$ for test datasets.

\section{\label{sec:level4}Model Construction}

A deep learning model utilizes a complex neural network for its architecture. A typical neural network has many layers and parameters with nonlinear functions to handle complex problems. The training data is first prepared, as in the previous section. Second, a specific model architecture is chosen. Model parameters in the architecture are then adjusted using the training data with a certain loss function and optimization algorithm. Finally, the performance of the trained model is tested through data unseen during the training process.

We used 1-dimensional convolutional layers and a dense layer at the end for the architecture. Convolutional neural networks have exceptionally performed well in various machine learning tasks \cite{ImageClassification, ImageNet, OverFeat, Segmentation, WaveNet, DCGAN}. Filters in a convolutional layer move around and are multiplied by the input feature data, producing the output data. The values, or weights, of filters are trained to find patterns in data. A dense layer at the end of the model sets the output size as 100, the size of waveforms.

\begin{table}[tp]
\caption{\label{tab:table1}Details of model structure. Batch normalization is located after every convolutional layer \cite{S.Ioffe2015}.}
\begin{tabular}{ccccc} \hline \hline 
&Layer&&Number of Filters (Units)& \\ \hline
 &Conv1D&&160&\\
 &Conv1D&&96&\\
 &MaxPooling1D&&&\\
 &Conv1D&&128&\\
 &Conv1D&&96&\\
 &Conv1D&&128&\\
 &MaxPooling1D&&&\\
 &Conv1D&&96&\\
 &Dense&&100&\\ \hline \hline
\end{tabular}
\end{table}

We used the Adam optimizer with a learning rate of 0.001 to optimize model parameters \cite{Adam}. Training a deep learning model requires a proper loss function. Mean squared error, a standard loss function for a regression problem, was used. It gives a measure of closeness between the true and predicted whole waveforms. Though our goal is to restore the whole particle waveform, the energy and timing information is most important in the TexAT data. The differences between the predictions and true waveforms on the peak heights and times were mainly monitored during the training process. We tuned hyperparameters of the model, such as the number of layers and filters, using the Keras Tuner framework and Hyperband algorithm \cite{KerasTuner, Hyperband}. The framework helped to find the optimized values for hyperparameters that have the best performance \cite{KerasTuner}. Details of the tuning algorithm Hyperband can be found in Ref. \cite{Hyperband}. Table~\ref{tab:table1} shows a summary of the tuned model structure.

\section{\label{sec:level5}Performance}

\subsection{\label{sec:a}Evaluation on the test dataset}

\begin{table}[tp]
\caption{\label{tab:table2}Summary of model performance on the test dataset. "Amplitude Loss" and "Timing Loss" show the errors of predictions on peak amplitude and timing. "Mean Squared Error" shows the errors of predictions on amplitudes of whole waveforms. See text for more details.}
\begin{tabular}{ccccccc} \hline \hline 
 &Type 1&&Type 2&&Total& \\ \hline
Amplitude Loss (\%)&5.0&&15.2&&10.1& \\
Timing Loss (Bin)&0.396&&1.245&&0.820& \\
Mean Squared Error&0.0009&&0.0011&&0.0010& \\ \hline \hline
\end{tabular}
\end{table}

\begin{figure*}[!t]
\includegraphics[width=7in]{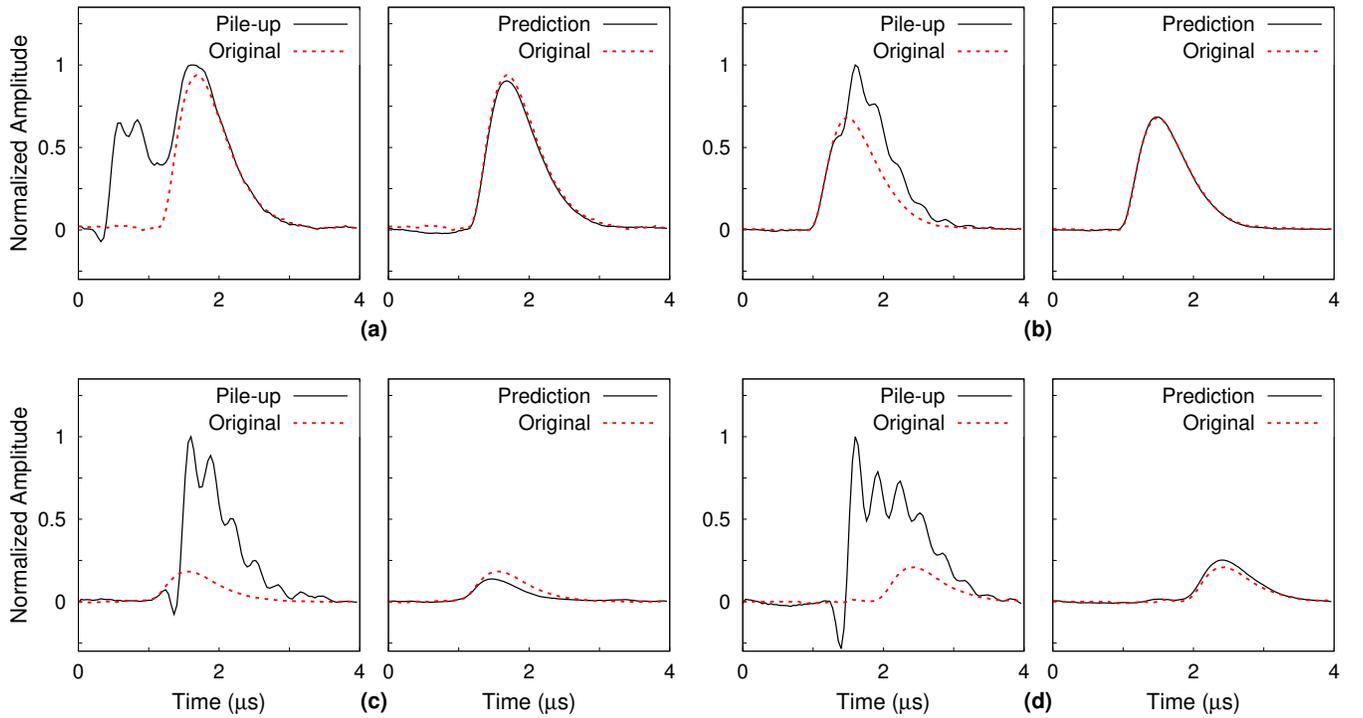}	
\caption{\label{fig:fig5}Examples of input pile-up, original (true), and predicted waveforms of the test dataset. The first ((a) and (b)) and second ((c) and (d)) rows are examples of Type 1 and 2 signals, respectively. See text for more details.}
\end{figure*}

\begin{figure*}[!t]
\includegraphics[width=7in]{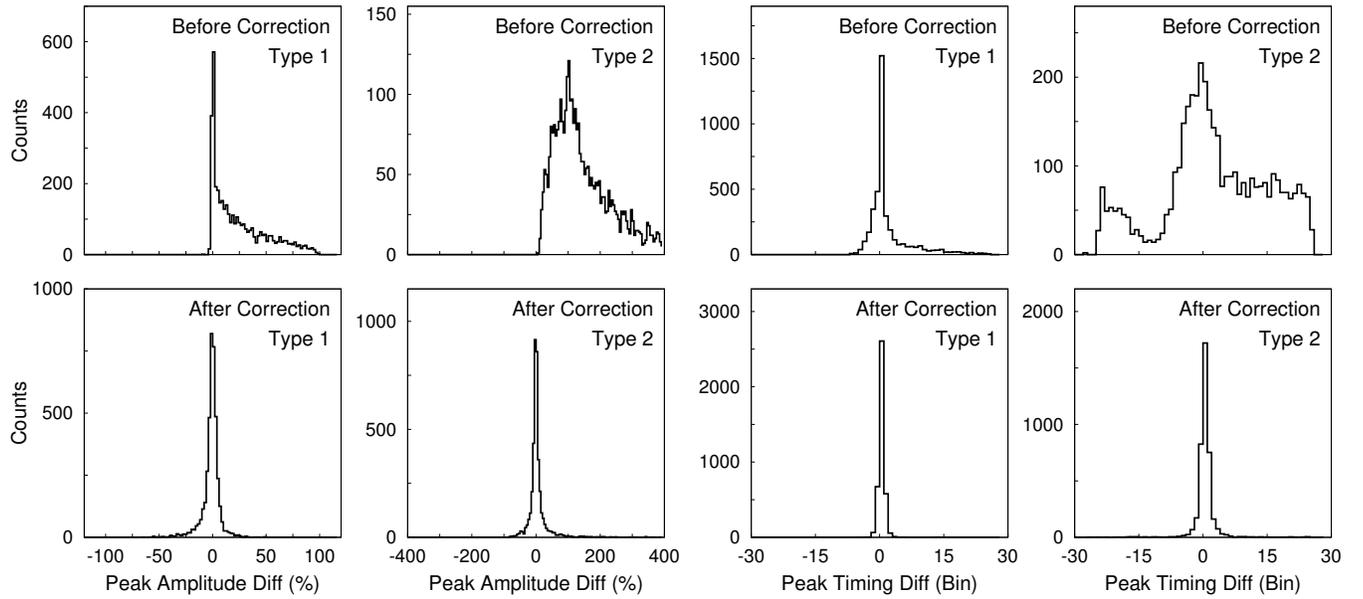}	
\caption{\label{fig:fig6}Histograms of peak amplitude and timing errors. Histograms in the first row are the errors without the correction using the deep learning model, whereas those in the second are the errors with the correction. See text for more details.}
\end{figure*}

We evaluated the trained model performance first by the test dataset. The model successfully reconstructed the original waveforms from the pile-up waveforms. Table~\ref{tab:table2} shows a summary of the evaluation. "Amplitude Loss" is defined as the average of $|\!\left(A_{\text{Prediction}} - A_{\text{True}}\right)\!/A_{\text{True}}|\times100$, where $A$ represents the peak amplitude. "Timing Loss" is the average of peak timing differences between model predictions and true waveforms. While these two metrics show the performance of peak information, "Mean Squared Error" gives the precision of predicted whole waveforms. The values of metrics in Table~\ref{tab:table2} demonstrate the high performance of the trained model.

\begin{figure*}[!t]
\includegraphics[width=7in]{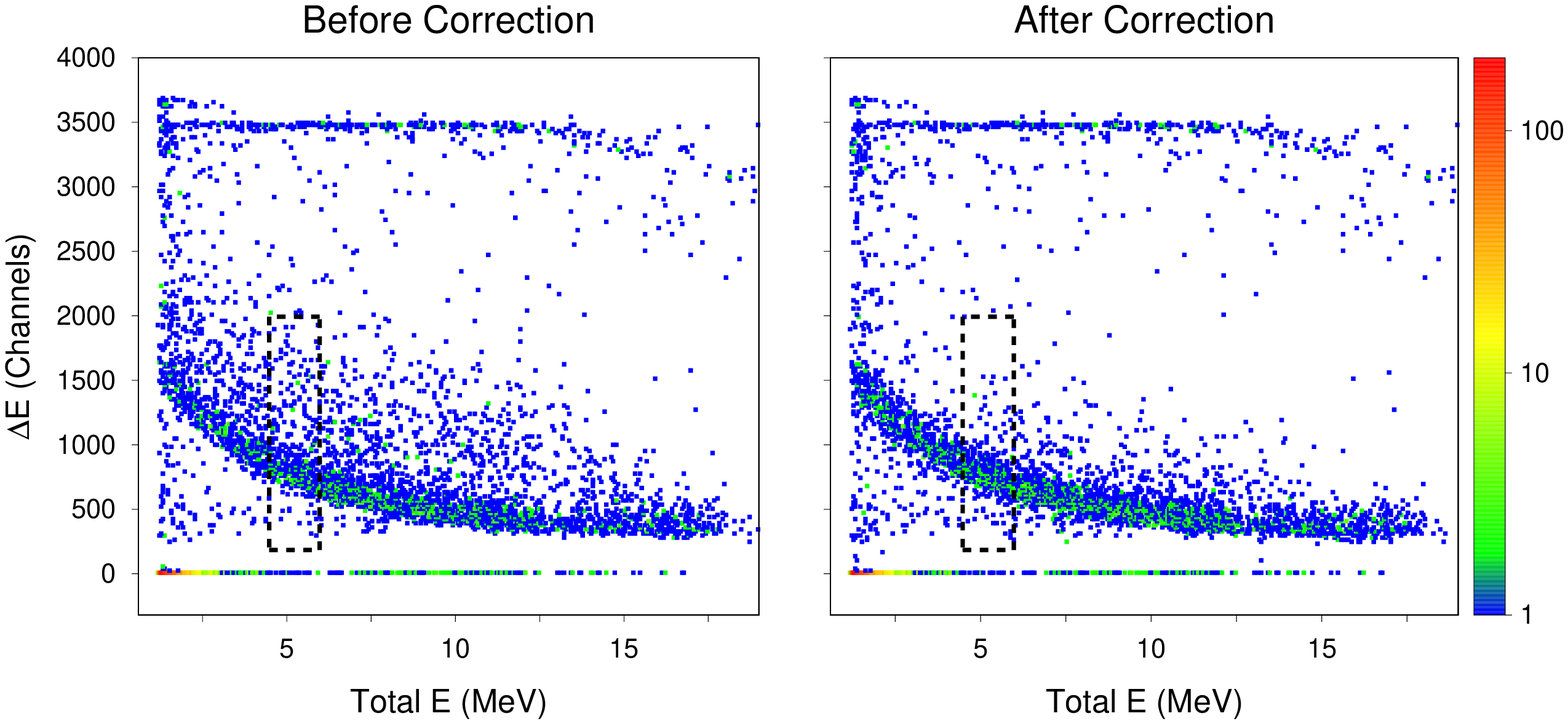}	
\caption{\label{fig:fig7}Comparison between original and new PID plots. The new PID plot is obtained using the correction to the pile-up signals.}
\end{figure*}

\begin{figure}[!t]
\includegraphics[width=3.3in]{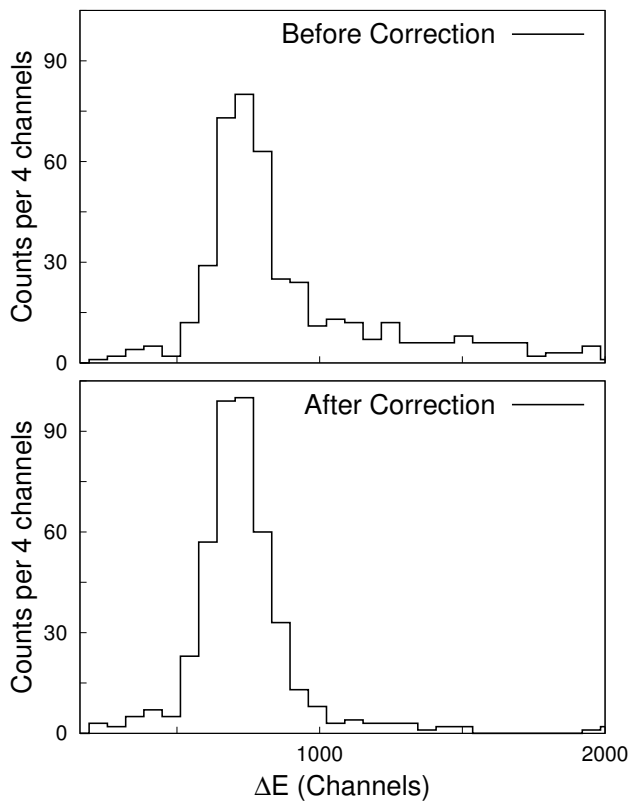}	
\caption{\label{fig:fig8}$\Delta E$ histograms of events inside the boxes in Fig.~\ref{fig:fig7}. See text for more details.}
\end{figure}

\begin{figure}[!t]
\includegraphics[width=3.3in]{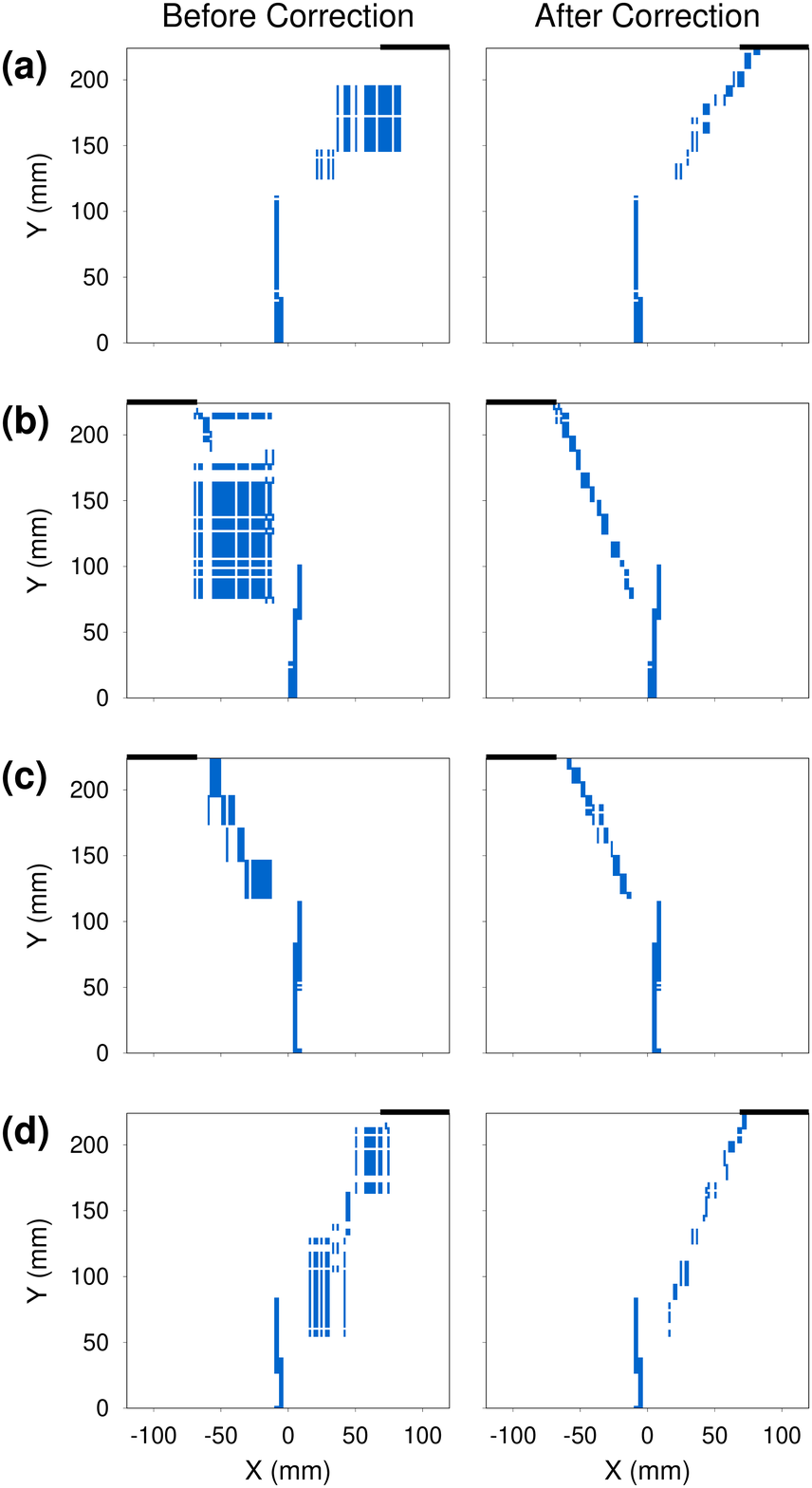}	
\caption{\label{fig:fig9}Reconstructed particle track images with and without the signal restorations. The black lines on top of each image present the approximate position of the fired silicon detector.}
\end{figure}

Fig.~\ref{fig:fig5} shows various examples of the pile-up, original, and predicted (restored) waveforms. The first row shows two examples of Type 1 signals. The left and right panels of each subplot are the pile-up and restored signals, respectively, along with the original signal. Even though the pile-up signals are complex, the model successfully restored the original signals. The second row shows two examples of Type 2 signals. While the overall model performance on the Type 2 signals is lower than on the Type 1 signals, the model still restored the small waveforms overlapped with large noise waveforms with high precision. Additionally, the model effectively works on various positions of original waveforms. Fig.~\ref{fig:fig5} (b) and (c) show cases where pile-up signals subside lower than the original signals due to negative noises. Fig.~\ref{fig:fig5} (d) shows a case where the original signal entirely resides in the noise signals.

Fig.~\ref{fig:fig6} shows comparisons between the errors of the pile-up and restored (corrected) signals. It presents histograms of the peak amplitude and timing errors before and after the correction on the test dataset using the model, which clearly shows improvements. The mean errors of peak amplitude decreased from 24 to 5 $\%$ for Type 1 and 259 to 15 $\%$ for Type 2. The mean errors of peak timing decreased from 2.7 to 0.4 bins for Type 1 and 9.7 to 1.2 bins for Type 2 (See Table~\ref{tab:table2}). To reconstruct particle tracks with the Micromegas, we match the peak timings of signals in the strips and chains normally within 40 ns (1 bin) (see Section~\ref{sec:level2}). If the timings are shifted by more than one bin because of the pile-up effects, track images will be distorted with low resolution. Before the correction, 43 and 78 $\%$ of Type 1 and 2 signals have peak timings shifted by more than one bin. After the correction by the model, they decreased to 4 and 18 $\%$. This implies that the correction can improve the track reconstruction (see Section~\ref{sec:b}).

\subsection{\label{sec:b}Evaluation on the real data}

The results on the test dataset only present the performance on the labeled and artificially added signals. We performed an additional test by applying the model to the real experimental data, where we do not know the true original signals. The model processed the signals classified as pile-up signals by the previous study of Kim \textit{et al.} \cite{C.Kim2023}. The pile-up signals were then replaced by the outputs of the model, the corrected signals. The energy and timing resolutions should increase if the model effectively works on the real data. We compared the new PID plot and track images with the original PID plot and track images to see such effects. Here, the original data is the data after removing the signals classified as the noise signals by the previous study \cite{C.Kim2023}.

Fig.~\ref{fig:fig7} shows the PID plots. The original (Before Correction) and new (After Correction) PID plots are presented in the left and right panels, respectively. $E$ is the calibrated energy loss measured in the solid-state detectors. $\Delta E$ is the energy loss per unit pad in the side region of the Micromegas detector (see Section~\ref{sec:level2}) \cite{J.Hooker2019, E.Koshchiy2020, C.Kim2023}. Therefore, the correction effect to the pile-up signals in the Micromegas will appear on the $\Delta E$ of the events. The plots show that the correction vividly enhances the energy resolution of the data to the pile-up signals. The pile-up signals normally have higher peak heights than the original signals. Because of this effect, $\Delta E$ of numerous events were certainly higher than the PID curve, which can be seen in the original PID plot. The new PID plot shows that these events moved back into the PID curve after the correction. Additionally, Fig.~\ref{fig:fig8} shows $\Delta E$ histograms of the events inside the dashed black boxes in Fig.~\ref{fig:fig7}. The histogram without the correction has a tail over the higher energy region because of the pile-up effects. The histogram with the correction does not have such a tail and exhibits a clearer Gaussian shape.

We use peak timings of signals in the Micromegas to reconstruct track images. The inaccurate timing information of the pile-up signals creates distorted particle paths. The improvement by the model on the timing information can be seen in the track reconstruction of the events that contain pile-up signals. Fig.~\ref{fig:fig9} shows examples of the original particle track images and corresponding improved images. The black line on the top of each image shows the approximate position of the Si detector fired by the particle. The original track images of Fig.~\ref{fig:fig9} (a) and (b) barely exhibit particle trajectories, which can increase uncertainties in track fittings. One also can raise the question of whether these are actual physical events. On the other hand, the corresponding track images with the corrected signals exhibit sharp particle trajectories pointing toward the triggered detectors. Additionally, Fig.~\ref{fig:fig9} (c) and (d) show that the thickness of the particle track can be reduced. These enhancements will present better resolution of the angular information, center of mass energy, and so on. The improved track images clearly show that the deep learning model gives effective timing corrections to pile-up signals.

\section{\label{sec:level6}Conclusion and Future Work}

Pile-up signals distort physics information such as energy and timing. Restoring original signals from pile-up signals will be very useful in experimental physics. The current study utilizes a deep learning model to reconstruct the original signals from the pile-up signals with noises. We showed that the model could restore the original signals accurately. The corrected heights of the pile-up signals effectively enhanced the energy resolution of the PID plot. The particle track reconstruction that uses the peak timings of signals was also notably improved. These achievements imply that the correction by the model will decrease the data analysis errors related to the energy, PID, and particle track. Practically, the denoising autoencoder with a deep classifier such as the one in the Ref. \cite{C.Kim2023} that classifies experimental signals into pile-up and non-pile-up signals will be greatly helpful.

The current method can also be used for any other types of noise. If the noises are completely random, the original unsupervised denoising autoencoder can be used. If the noises have a certain manner, the same procedure of this study can be performed by collecting such noises and clean signals. Deep learning models can also handle other tasks, such as separating multiple desired signals piled up from each other. In such a case, one simply needs to make a model that outputs multiple separate waveforms. Additionally, one can build a model that smooths out signals with constant background fluctuations. The implementation of deep learning in signal processing can improve the quality of signals and, therefore, physics analysis.

\section*{Acknowledgements}

This work was supported by the National Research Foundation of Korea (NRF) grants funded by the Korea government (MSIT) (Grants No. 2016R1A5A1013277 and No. 2020R1A2C1005981). This work was also supported in part by the Institute for Basic Science (Grant No. IBS-R031-D1) and the U.S. Department of Energy under Award No. DE-FG03-93ER40773. Computational works for this research were performed on the data analysis hub, Olaf in the IBS Research Solution Center.

\bibliography{Denoiser}{}

\begin{thebibliography}{10}
\expandafter\ifx\csname url\endcsname\relax
  \def\url#1{\texttt{#1}}\fi
\expandafter\ifx\csname urlprefix\endcsname\relax\def\urlprefix{URL }\fi
\expandafter\ifx\csname href\endcsname\relax
  \def\href#1#2{#2} \def\path#1{#1}\fi

\bibitem{Z.Marshall2014}
Z.~{Marshall}, {Atlas Collaboration}, {Simulation of Pile-up in the ATLAS
  Experiment}, in: Journal of Physics Conference Series, Vol. 513 of Journal of
  Physics Conference Series, 2014, p. 022024.
\newblock \href {https://doi.org/10.1088/1742-6596/513/2/022024}
  {\path{doi:10.1088/1742-6596/513/2/022024}}.

\bibitem{M.Hammad2019}
M.~Hammad, H.~Kasban, R.~Fikry, M.~I. Dessoky, O.~Zahran, S.~M. Elaraby, F.~E.
  {Abd El-Samie},
  {Pile-up
  correction algorithm for high count rate gamma ray spectroscopy}, Applied
  Radiation and Isotopes 151 (2019) 196--206.
\newblock \href
  {https://doi.org/https://doi.org/10.1016/j.apradiso.2019.06.003}
  {\path{doi:https://doi.org/10.1016/j.apradiso.2019.06.003}}.


\bibitem{M.Mohammadian-Behbahani2020}
M.-R. {Mohammadian-Behbahani}, S.~{Saramad}, {A comparison study of the pile-up
  correction algorithms}, Nuclear Instruments and Methods in Physics Research A
  951 (2020) 163013.
\newblock \href {https://doi.org/10.1016/j.nima.2019.163013}
  {\path{doi:10.1016/j.nima.2019.163013}}.

\bibitem{ImageClassification}
W.~Rawat, Z.~Wang, Deep convolutional neural networks for image classification:
  A comprehensive review, Neural Computation 29~(9) (2017) 2352--2449.
\newblock \href {https://doi.org/https://doi.org/10.1162/neco\_a\_00990}
  {\path{doi:https://doi.org/10.1162/neco\_a\_00990}}.

\bibitem{B.Rem2019}
B.~S. {Rem}, N.~{K{\"a}ming}, M.~{Tarnowski}, L.~{Asteria}, N.~{Fl{\"a}schner},
  C.~{Becker}, K.~{Sengstock}, C.~{Weitenberg}, {Identifying quantum phase
  transitions using artificial neural networks on experimental data}, Nature
  Physics 15~(9) (2019) 917--920.
\newblock \href {http://arxiv.org/abs/1809.05519} {\path{arXiv:1809.05519}},
  \href {https://doi.org/10.1038/s41567-019-0554-0}
  {\path{doi:10.1038/s41567-019-0554-0}}.

\bibitem{H.Gabbard2022}
H.~{Gabbard}, C.~{Messenger}, I.~S. {Heng}, F.~{Tonolini}, R.~{Murray-Smith},
  {Bayesian parameter estimation using conditional variational autoencoders for
  gravitational-wave astronomy}, Nature Physics 18~(1) (2022) 112--117.
\newblock \href {http://arxiv.org/abs/1909.06296} {\path{arXiv:1909.06296}},
  \href {https://doi.org/10.1038/s41567-021-01425-7}
  {\path{doi:10.1038/s41567-021-01425-7}}.

\bibitem{A.Senior2020}
A.~W. {Senior}, R.~{Evans}, J.~{Jumper}, J.~{Kirkpatrick}, L.~{Sifre},
  T.~{Green}, C.~{Qin}, A.~{{\v{Z}}{\'\i}dek}, A.~W.~R. {Nelson},
  A.~{Bridgland}, H.~{Penedones}, S.~{Petersen}, K.~{Simonyan}, S.~{Crossan},
  P.~{Kohli}, D.~T. {Jones}, D.~{Silver}, K.~{Kavukcuoglu}, D.~{Hassabis},
  {Improved protein structure prediction using potentials from deep learning},
  Nature 577~(7792) (2020) 706--710.
\newblock \href {https://doi.org/10.1038/s41586-019-1923-7}
  {\path{doi:10.1038/s41586-019-1923-7}}.

\bibitem{J.Jumper2021}
J.~{Jumper}, R.~{Evans}, A.~{Pritzel}, T.~{Green}, M.~{Figurnov},
  O.~{Ronneberger}, K.~{Tunyasuvunakool}, R.~{Bates}, A.~{{\v{Z}}{\'\i}dek},
  A.~{Potapenko}, A.~{Bridgland}, C.~{Meyer}, S.~A.~A. {Kohl}, A.~J. {Ballard},
  A.~{Cowie}, B.~{Romera-Paredes}, S.~{Nikolov}, R.~{Jain}, J.~{Adler},
  T.~{Back}, S.~{Petersen}, D.~{Reiman}, E.~{Clancy}, M.~{Zielinski},
  M.~{Steinegger}, M.~{Pacholska}, T.~{Berghammer}, S.~{Bodenstein},
  D.~{Silver}, O.~{Vinyals}, A.~W. {Senior}, K.~{Kavukcuoglu}, P.~{Kohli},
  D.~{Hassabis}, {Highly accurate protein structure prediction with AlphaFold},
  Nature 596~(7873) (2021) 583--589.
\newblock \href {https://doi.org/10.1038/s41586-021-03819-2}
  {\path{doi:10.1038/s41586-021-03819-2}}.

\bibitem{M.Raissi2019}
M.~{Raissi}, P.~{Perdikaris}, G.~E. {Karniadakis}, {Physics-informed neural
  networks: A deep learning framework for solving forward and inverse problems
  involving nonlinear partial differential equations}, Journal of Computational
  Physics 378 (2019) 686--707.
\newblock \href {https://doi.org/10.1016/j.jcp.2018.10.045}
  {\path{doi:10.1016/j.jcp.2018.10.045}}.

\bibitem{C.Kim2023}
C.~Kim, S.~Ahn, K.~Chae, J.~Hooker, G.~Rogachev,
  \href{https://www.sciencedirect.com/science/article/pii/S0168900223000153}{Noise
  signal identification in time projection chamber data using deep learning
  model}, Nuclear Instruments and Methods in Physics Research Section A:
  Accelerators, Spectrometers, Detectors and Associated Equipment 1048 (2023)
  168025.
\newblock \href {https://doi.org/https://doi.org/10.1016/j.nima.2023.168025}
  {\path{doi:https://doi.org/10.1016/j.nima.2023.168025}}.


\bibitem{C.Zhang2016}
C.-X. {Zhang}, S.-T. {Lin}, J.-L. {Zhao}, X.-Z. {Yu}, L.~{Wang}, J.-J. {Zhu},
  H.-Y. {Xing}, {Discrimination of neutrons and {\ensuremath{\gamma}}-rays in
  liquid scintillator based on Elman neural network}, Chinese Physics C 40~(8)
  (2016) 086204.
\newblock \href {http://arxiv.org/abs/1509.06259} {\path{arXiv:1509.06259}},
  \href {https://doi.org/10.1088/1674-1137/40/8/086204}
  {\path{doi:10.1088/1674-1137/40/8/086204}}.

\bibitem{C.Yang2017}
C.~{Yang}, C.~{Feng}, W.~{Dong}, D.~{Jiang}, Z.~{Shen}, S.~{Liu}, Q.~{An},
  {Alpha-Gamma Discrimination in BaF2 Using FPGA-Based Feedforward Neural
  Network}, IEEE Transactions on Nuclear Science 64~(6) (2017) 1350--1356.
\newblock \href {https://doi.org/10.1109/TNS.2017.2691729}
  {\path{doi:10.1109/TNS.2017.2691729}}.

\bibitem{P.Holl2019}
P.~{Holl}, L.~{Hauertmann}, B.~{Majorovits}, O.~{Schulz}, M.~{Schuster}, A.~J.
  {Zsigmond}, {Deep learning based pulse shape discrimination for germanium
  detectors}, European Physical Journal C 79~(6) (2019) 450.
\newblock \href {http://arxiv.org/abs/1903.01462} {\path{arXiv:1903.01462}},
  \href {https://doi.org/10.1140/epjc/s10052-019-6869-2}
  {\path{doi:10.1140/epjc/s10052-019-6869-2}}.

\bibitem{L.Garcia2021}
L.~G. Garcia, R.~S. Molina, M.~L. Crespo, S.~Carrato, G.~Ramponi, A.~Cicuttin,
  I.~R. Morales, H.~Perez, Muon–electron pulse shape discrimination for water
  cherenkov detectors based on fpga/soc, Electronics 10~(3) (2021).
\newblock \href {https://doi.org/10.3390/electronics10030224}
  {\path{doi:10.3390/electronics10030224}}.

\bibitem{M.Kafaee2009}
M.~{Kafaee}, S.~{Saramad}, {Pile-up correction by Genetic Algorithm and
  Artificial Neural Network}, Nuclear Instruments and Methods in Physics
  Research A 607~(3) (2009) 652--658.
\newblock \href {https://doi.org/10.1016/j.nima.2009.06.033}
  {\path{doi:10.1016/j.nima.2009.06.033}}.

\bibitem{P.Vincent2008}
P.~Vincent, H.~Larochelle, Y.~Bengio, P.-A. Manzagol,
  \href{https://doi.org/10.1145/1390156.1390294}{Extracting and composing
  robust features with denoising autoencoders}, in: Proceedings of the 25th
  International Conference on Machine Learning, ICML '08, Association for
  Computing Machinery, New York, NY, USA, 2008, p. 1096–1103.
\newblock \href {https://doi.org/10.1145/1390156.1390294}
  {\path{doi:10.1145/1390156.1390294}}.
\newline\urlprefix\url{https://doi.org/10.1145/1390156.1390294}

\bibitem{P.Vincent2010}
P.~Vincent, H.~Larochelle, I.~Lajoie, Y.~Bengio, P.-A. Manzagol,
  \href{http://jmlr.org/papers/v11/vincent10a.html}{Stacked denoising
  autoencoders: Learning useful representations in a deep network with a local
  denoising criterion}, Journal of Machine Learning Research 11~(110) (2010)
  3371--3408.
\newline\urlprefix\url{http://jmlr.org/papers/v11/vincent10a.html}

\bibitem{E.Koshchiy2020}
E.~{Koshchiy}, G.~V. {Rogachev}, E.~{Pollacco}, S.~{Ahn}, E.~{Uberseder},
  J.~{Hooker}, J.~{Bishop}, E.~{Aboud}, M.~{Barbui}, V.~Z. {Goldberg},
  C.~{Hunt}, H.~{Jayatissa}, C.~{Magana}, R.~{O'Dwyer}, B.~T. {Roeder},
  A.~{Saastamoinen}, S.~{Upadhyayula}, {Texas Active Target (TexAT) detector
  for experiments with rare isotope beams}, Nuclear Instruments and Methods in
  Physics Research A 957 (2020) 163398.
\newblock \href {http://arxiv.org/abs/1906.07845} {\path{arXiv:1906.07845}},
  \href {https://doi.org/10.1016/j.nima.2020.163398}
  {\path{doi:10.1016/j.nima.2020.163398}}.

\bibitem{J.Hooker2019}
J.~{Hooker}, G.~V. {Rogachev}, E.~{Koshchiy}, S.~{Ahn}, M.~{Barbui}, V.~Z.
  {Goldberg}, C.~{Hunt}, H.~{Jayatissa}, E.~C. {Pollacco}, B.~T. {Roeder},
  A.~{Saastamoinen}, S.~{Upadhyayula}, {Structure of $^{9}$C through proton
  resonance scattering with the Texas Active Target detector}, Physical Review
  C 100~(5) (2019) 054618.
\newblock \href {http://arxiv.org/abs/1903.01402} {\path{arXiv:1903.01402}},
  \href {https://doi.org/10.1103/PhysRevC.100.054618}
  {\path{doi:10.1103/PhysRevC.100.054618}}.

\bibitem{J.Bishop2020}
J.~{Bishop}, G.~V. {Rogachev}, S.~{Ahn}, E.~{Aboud}, M.~{Barbui}, A.~{Bosh},
  C.~{Hunt}, H.~{Jayatissa}, E.~{Koshchiy}, R.~{Malecek}, S.~T. {Marley}, E.~C.
  {Pollacco}, C.~D. {Pruitt}, B.~T. {Roeder}, A.~{Saastamoinen}, L.~G.
  {Sobotka}, S.~{Upadhyayula}, {Almost medium-free measurement of the Hoyle
  state direct-decay component with a TPC}, Physical Review C 102~(4) (2020)
  041303.
\newblock \href {http://arxiv.org/abs/2012.08437} {\path{arXiv:2012.08437}},
  \href {https://doi.org/10.1103/PhysRevC.102.041303}
  {\path{doi:10.1103/PhysRevC.102.041303}}.

\bibitem{J.Zamora2021}
J.~C. {Zamora}, V.~{Guimaraes}, G.~V. {Rogachev}, S.~{Ahn}, J.~{Lubian}, E.~N.
  {Cardozo}, E.~{Aboud}, M.~{Assuncao}, M.~{Barbui}, J.~{Bishop}, A.~{Bosh},
  J.~{Hooker}, C.~{Hunt}, H.~{Jayatissa}, E.~{Koshchiy}, S.~{Lukyanov},
  R.~{O'Dwyer}, Y.~{Penionzhkevich}, B.~T. {Roeder}, A.~{Saastamoinen},
  S.~{Upadhyayula}, {Direct fusion measurement of the $^{8}$B proton-halo
  nucleus at near-barrier energies}, Physics Letters B 816 (2021) 136256.
\newblock \href {https://doi.org/10.1016/j.physletb.2021.136256}
  {\path{doi:10.1016/j.physletb.2021.136256}}.

\bibitem{J.Bishop2022}
J.~{Bishop}, C.~E. {Parker}, G.~V. {Rogachev}, S.~{Ahn}, E.~{Koshchiy},
  K.~{Brandenburg}, C.~R. {Brune}, R.~J. {Charity}, J.~{Derkin}, N.~{Dronchi},
  G.~{Hamad}, Y.~{Jones-Alberty}, T.~{Kokalova}, T.~N. {Massey}, Z.~{Meisel},
  E.~V. {Ohstrom}, S.~N. {Paneru}, E.~C. {Pollacco}, M.~{Saxena}, N.~{Singh},
  R.~{Smith}, L.~G. {Sobotka}, D.~{Soltesz}, S.~K. {Subedi}, A.~V. {Voinov},
  J.~{Warren}, C.~{Wheldon}, {Neutron-upscattering enhancement of the
  triple-alpha process}, Nature Communications 13 (2022) 2151.
\newblock \href {https://doi.org/10.1038/s41467-022-29848-7}
  {\path{doi:10.1038/s41467-022-29848-7}}.

\bibitem{ImageNet}
A.~Krizhevsky, I.~Sutskever, G.~E. Hinton, Imagenet classification with deep
  convolutional neural networks, in: F.~Pereira, C.~J.~C. Burges, L.~Bottou,
  K.~Q. Weinberger (Eds.), Advances in Neural Information Processing Systems,
  Vol.~25, Curran Associates, Inc., 2012.

\bibitem{OverFeat}
P.~{Sermanet}, D.~{Eigen}, X.~{Zhang}, M.~{Mathieu}, R.~{Fergus}, Y.~{LeCun},
  {OverFeat: Integrated Recognition, Localization and Detection using
  Convolutional Networks}, arXiv e-prints (2013) arXiv:1312.6229.

\bibitem{Segmentation}
J.~{Long}, E.~{Shelhamer}, T.~{Darrell}, {Fully Convolutional Networks for
  Semantic Segmentation}, arXiv e-prints (2014) arXiv:1411.4038.

\bibitem{WaveNet}
A.~{van den Oord}, S.~{Dieleman}, H.~{Zen}, K.~{Simonyan}, O.~{Vinyals},
  A.~{Graves}, N.~{Kalchbrenner}, A.~{Senior}, K.~{Kavukcuoglu}, {WaveNet: A
  Generative Model for Raw Audio}, arXiv e-prints (2016) arXiv:1609.03499.

\bibitem{DCGAN}
A.~{Radford}, L.~{Metz}, S.~{Chintala}, {Unsupervised Representation Learning
  with Deep Convolutional Generative Adversarial Networks}, arXiv e-prints
  (2015) arXiv:1511.06434.

\bibitem{S.Ioffe2015}
S.~{Ioffe}, C.~{Szegedy}, {Batch Normalization: Accelerating Deep Network
  Training by Reducing Internal Covariate Shift}, arXiv e-prints (2015)
  arXiv:1502.03167\href {http://arxiv.org/abs/1502.03167}
  {\path{arXiv:1502.03167}}.

\bibitem{Adam}
D.~P. {Kingma}, J.~{Ba}, {Adam: A Method for Stochastic Optimization}, arXiv
  e-prints (2014) arXiv:1412.6980.

\bibitem{KerasTuner}
T.~O'Malley, E.~Bursztein, J.~Long, F.~Chollet, H.~Jin, L.~Invernizzi, et~al.,
  Kerastuner, \url{https://github.com/keras-team/keras-tuner} (2019).

\bibitem{Hyperband}
L.~{Li}, K.~{Jamieson}, G.~{DeSalvo}, A.~{Rostamizadeh}, A.~{Talwalkar},
  {Hyperband: A Novel Bandit-Based Approach to Hyperparameter Optimization},
  arXiv e-prints (2016) arXiv:1603.06560.

\end{thebibliography}
\bibliographystyle{elsarticle-num}
\end{document}